\documentclass[pre,aps,showpacs,showkeys,12pt]{revtex4}
\usepackage{wrapfig}
\usepackage{graphicx}
\usepackage[english,russian]{babel}
\usepackage[cp1251]{inputenc}
\usepackage{amsmath}

\begin{document}

\title{The weighed average geodetic of distributions of probabilities in
the statistical physics}

\author{V.V. Ryazanov}

\affiliation{Institute for Nuclear Research,  Kiev,  pr. Nauki,
 47, Ukraine ; \quad e-mail:vryazan@kinr.kiev.ua}

\begin{abstract}

The results received in works [Centsov N.N. [N.N. Chentsov],
Statistical decision rules and
 optimal inference, 1982 Amer. Math. Soc.   (Translated from
 Russian); Morozova, E. A., Chentsov, N. N. Natural geometry of families
 of probability laws. 1991 Probability theory, 8, 133--265, 270--274, 276 (in Russian)] for
statistical distributions at studying algebra of decision rules
and natural geometry generated by her, are applied to estimations
of the nonequilibrium statistical operator and superstatistics.
Expressions for the nonequilibrium statistical operator and
superstatistics are received as special cases of the weighed
average geodetic of distributions of probabilities.
\end{abstract}



\maketitle

%{\small $\dag$ e-mail: vryazan@kinr.kiev.ua }

\section{Introduction}

 In works \cite {Chentsov, Moroz} the
differential geometry of varieties of probabilistic measures which
gives a natural language as the description of statistical model -
to the a priori information on statistical experiment, and
constructions of optimum methods of processing of such experiment
is investigated. It is possible to interpret many results of works
\cite {Chentsov, Moroz} in terms of statistical physics. It
concerns to exponent families of distributions, to "spread"   of
singular measures on all convex bearer, to problems of projecting,
inequalities of the information, and other features of behaviour
of the probabilistic distributions studied in \cite {Chentsov,
Moroz}.   In the present work communication of the weighed average
geodetic of continuous family of probabilistic laws \cite
{Chentsov, Moroz} with the nonequilibrium statistical operator
(NSO) \cite {Zub71, Zub80, Zub99} and with superstatistics \cite
{Beck, BeckCohen} is traced.

\section{The weighed average geodetic of distributions of probabilities}

Following \cite {Chentsov}, we shall describe a class of
probabilistic families for which in \cite {Chentsov, Moroz} the
notion of the weighed average geodetic is defined. We consider,
that the smooth family $ \overrightarrow {\Phi} $ of probabilistic
laws can be described by means of the unique open map $(\Theta,
\varphi), \overrightarrow {x} = \varphi (P)$, or return reflection
$\Theta\stackrel{\Psi}{\rightarrow} Caph(\Omega,
\overrightarrow{S}, \overrightarrow{Z})$, where $ \Omega $ is
space of all elementary outcomes $ \omega $ (of experiment),
$\overrightarrow {S}$ is some $ \sigma $-algebra of its subsets
named also by events. For each measure $ \mu {\{\cdot \}} $ on
measurable space $ (\Omega, \overrightarrow {S}) $ all sets zero
measures (zero-sets) form an ideal $ \overrightarrow {Z} =
\overrightarrow {Z _ {\mu}} $ of algebra $ \overrightarrow {S} $.
Set of all probabilistic measures cancelled on an ideal $
\overrightarrow {Z} $ and only on $ \overrightarrow {Z} $, is
designated through $Caph (\Omega, \overrightarrow {S},
\overrightarrow {Z}) $. They form undertotality mutually
absolutely continuous distributions on $ (\Omega, \overrightarrow
{S}) $.  If two measures $ \mu $ and $ \nu $ have the general
ideal zero-sets them name mutually absolutely continuous (or
quasi-equivalent).  If $ \overrightarrow {Z _ {\mu}} \subseteq
\overrightarrow {Z _ {\nu}} $ speak, that $ \mu $ dominates $ \nu
$, and write down it: $ \mu \gg \nu $.
 We name \cite {Chentsov} the
$m$-dimensional open map of set $M$ univocal reflection $ \varphi
$ of subset $ \Theta \subseteq M $ on the coherent open area of $m
$-dimensional Euclid space $R^{m}$. Coordinates $x^{(1)}(P)...,
x^{(m)}(P)$ of point $\varphi(P)$ thus refer to as local
coordinates of a point $P \in M$ on a considered map $(\Theta,
\varphi)$. Arising in variety $Caph (\Omega, \overrightarrow {S},
\overrightarrow {Z})$ the surface $\{P_{\overrightarrow {x}},
\overrightarrow {x}\in\Theta\}$ has no self-crossings, i.e.
reflection $\Psi = \varphi^{-1}$ biunique. The described families
refer to in \cite {Chentsov} simple.

The simple family of distributions of probabilities
$\{P_{\overrightarrow{x}},\overrightarrow{x}\in\Theta\} \subset
Caph (\Omega, \overrightarrow {S}, \overrightarrow {Z})$ refers to
smooth \cite {Chentsov}, when

$1^{\circ}$ there is a coordinated variant of densities $p
(\omega; \overrightarrow {x})$ on the fixed measure $\mu \in Conh
(\Omega, \overrightarrow {S}, \overrightarrow {Z})$ such, that at
everyone $ \omega \in \Omega $ density $p (\omega; \overrightarrow
{x})$ there is three times differentiated positive function of
argument $(x _ {1}..., x _ {n}) = \overrightarrow {x} \in \Theta$
($Conh (\Omega, \overrightarrow {S}, \overrightarrow {Z})$ is set
of all non-negative mutually absolute continuous measures on
$(\Omega, \overrightarrow {S})$, addressing in zero on
$\overrightarrow {Z}$-sets and only on them);

$2^{\circ}$ at everyone $ \overrightarrow {x} \in \Theta $ partial
derivatives $p^{\prime}_{j} (\omega; \overrightarrow {x}) =
\partial p (\omega; \overrightarrow {x}) / \partial x _ {j} $
of density, $j=1..., n $, are linearly independent on $ \Omega $
even if to neglect their values on any $ \overrightarrow {Z}
$-set;

$3^{\circ}$ for everyone $\theta \in \Theta$ there will be a
special vicinity $O _ {\theta}$, in which derivatives
$p{\prime}_{j}(\omega; \overrightarrow {x})$ suppose a majorant
$g{(\theta)}(\omega)=(dG{(\theta)}/d\mu)(\omega)$:
\begin{equation}
\left.
\begin{array}{l}
 p^{\prime}_{j}(\omega; \overrightarrow{x})\leq
 g^{(\theta)}(\omega), \quad P^{\prime}_{j}\{\cdot \mid
 \overrightarrow{x}\} \leq G^{(\theta)}\{\cdot\},
 \quad  \forall{\overrightarrow{x}} \in
 O_{\theta}, \quad \forall\omega \in
 \Omega,
 \\
 M_{\overrightarrow{x}}[g^{\theta}(\omega)/p(\omega;
 \overrightarrow{x})]^{2}\leq L_{\theta}^{2} <
 \infty, \quad \forall{\overrightarrow{x}} \in
 O_{\theta}
 \end{array}
 \right\};
\label{g}
\end{equation}

$4^{\circ}$ for everyone $\theta \in \Theta$ in the specified
special vicinity $O _ {\theta}$ all partial derivatives of
likelihood function $\ln p(\omega;\overrightarrow{x})$ up to the
third order inclusive suppose a majorant

\begin{equation}
|\frac{\partial^{|\overrightarrow{k}|}\ln
p(\omega;\overrightarrow{x})}{\partial{\overrightarrow{x}^{\overrightarrow{k}}}}|\leq
h^{(\theta)}(\omega), \quad \forall \overrightarrow{x}\in
O_{\theta}, \quad |\overrightarrow{k}|=1,2,3,\,
 \label{consh}
\end{equation}

where
$\overrightarrow{k}=(k_{1},...,k_{n}), \quad
|\overrightarrow{k}|=k_{1}+...+k_{n}, \quad
\partial{\overrightarrow{x}^{\overrightarrow{k}}}=
\partial{x_{1}^{k_{1}}}...\partial{x_{n}^{k_{n}}}$, and

\begin{equation}
M_{\overrightarrow{x}}[h^{\theta}(\omega)]^{4}\leq H_{\theta}^{4}<
\infty, \quad \forall\overrightarrow{x} \in
 O_{\theta} .\
\label{constH}
\end{equation}
The constant $L _{\theta}^{2}$ from (\ref {g}) is defined through
estimations of the second derivatives.

Below us designations will be
necessary

\begin{equation}
\frac{\partial \ln p(\omega ; \overrightarrow{x})}{\partial x_j}=
r^{j}(\omega; \overrightarrow{x}), \qquad \frac{\partial^{2}{\ln
p(\omega; \overrightarrow{x})}}{\partial x_j
\partial x_k}= r^{jk}(\omega;
\overrightarrow{x})\, ,
 \label{r}
\end{equation}
and following consequence from a lemma 27.5 \cite {Chentsov}:

For smooth family it is identical

\begin{equation}
M_{\overrightarrow{x}}r^{j}(\omega; \overrightarrow{x})=0;
\label{M}
\end{equation}

\begin{equation}
-M_{\overrightarrow{x}}r^{jk}(\omega;\overrightarrow{x})=M_{\overrightarrow{x}}r^{j}(\omega;
\overrightarrow{x})r^{k}(\omega;
\overrightarrow{x})=\overrightarrow{\omega}^{jk}(\overrightarrow{x}),\,
 \label{M1}
\end{equation}
where $M_{\overrightarrow{x}}r^{jk}(\omega;\overrightarrow{x})=
\int_{\Omega}r^{jk}(\omega;\overrightarrow{x})p(\omega;\overrightarrow{x})\mu\{d\omega\}$
is averaging,
 $\overrightarrow {\omega}_{jk}(\theta)=M_{\theta}r^{j}(\omega;\theta)r^{k}(\omega;\theta)$ is
Fisher's information matrix. Alongside with initial
parametrization of family we shall consider also its linear
reparametrization. When in new system of coordinates Fisher's
information matrix in a point $\theta$ will be single such system
of coordinates refers to in \cite {Chentsov} $\theta$-local. In
\cite {Chentsov} it is considered and $\theta$-local distance
between laws $P _ {\overrightarrow {x}}$ and $P_{\overrightarrow
{\tau}}$:

\begin{equation}
||\overrightarrow{x}-\overrightarrow{\tau}||^{2}_{\theta}=\sum_{j,k}(x_{j}-\tau_{j})
(x_{k}-\tau_{k})\overrightarrow{\omega}^{jk}(\theta).\
\label{matrw}
\end{equation}

Let's enter now according to \cite {Chentsov} concept of the
weighed average geodetic of continuous family of probabilities
laws. Let the family
$\overrightarrow{\Phi}=\{P_{\overrightarrow{x}},\overrightarrow{x}\in
C\}$ distributions on $(\Omega, \overrightarrow {S})$, depending
on vector parameter $\overrightarrow{x}$ with compact set $C$
values of parameter, is set by family of coordinated strictly
positive densities $p (\omega; \overrightarrow {x})$ concerning a
measure $R$, continuous on $\overrightarrow {x}$ at everyone
$\omega \in \Omega$. Let, further, $\alpha\{\cdot\}$ is any
probabilities Borel's a measure on $C$. In \cite {Chentsov}
weighed (with a weight measure $\alpha$) average geodetic of laws
of family $\overrightarrow{\Phi}$ to refers distribution of
probabilities $U_{\alpha}$ with the logarithm of density

\begin{equation}
\ln u_{\alpha}(\omega)=\int_{C}\ln
p(\omega;\overrightarrow{x})\alpha\{d\overrightarrow{x}\}-H[\alpha],\,
\label{geod}
\end{equation}
where $H [\alpha]$ is the logarithm of a normalizing divider

\begin{equation}
expH[\alpha]=\int_{\Omega}\exp[\int_{C}\ln
p(\omega;\overrightarrow{x})\alpha\{d\overrightarrow{x}\}]R\{d\omega\}
,\, \label{norm}
\end{equation}
if only last integral is finite. We shall speak otherwise, that
the specified average does not exist.

In \cite {Chentsov} the set $\gamma \subset Caph(\Omega,
\overrightarrow{S}, \overrightarrow{Z})$ of distributions of
probabilities $P_{s}\{\cdot\}$ of exponent or geodetic family
(finite number of measurements) with canonical affine parameter
$\overrightarrow {s} = (s_{1}.., .s_{n})$ is entered also and with
family of density

\begin{equation}
\frac{dP_{s}}{d\mu}(\omega)=p(\omega;\overrightarrow{s})=p_{0}(\omega)
exp[\sum_{j}s^{j}q_{j}(\omega)-\Psi(s)] ,\, \label{exp}
\end{equation}
where $\overrightarrow {q} = (q _ {1} (\omega)..., q _ {n}
(\omega))$ is directing sufficient statistics \cite {Chentsov},
$\mu\{\cdot\}$ is the fixed dominating measure, and

\begin{equation}
exp[\Psi(s)]=\int_{\Omega} exp[\sum_{j}s^{j}q_{j}(\omega)]
p_{0}(\omega)\mu\{d\omega\}  \label{norm1}
\end{equation}
is a normalizing divider. It is supposed, that the parameter
$\overrightarrow{s}$ of distribution run all values at which the
normalizing divider is finite, i.e. $\gamma$ is the maximal family
of distributions, representable at the some $s^{1}..., s^{n}$ in
the form of (\ref {exp}). In \cite {Chentsov, Moroz} it is shown,
that the family of densities in (\ref {exp}) - (\ref {norm1}) is
"trajectory"   of $n$-dimensional subgroup of group of
translations of variety $Caph (\Omega, \overrightarrow {S},
\overrightarrow {Z})$ of distributions of probabilities.
Distributions (\ref {exp}) are included into wider class
exponential families with density of a kind

\begin{equation}
p(\omega;\overrightarrow{\theta})=p_{0}(\omega)
exp[\sum_{j}s^{j}(\overrightarrow{\theta})q_{j}(\omega)-\Psi(\overrightarrow{s}(\overrightarrow{\theta}))]
,\, \label{exp1}
\end{equation}
where $\overrightarrow{\theta}=(\theta_{1},...,\theta_{m})\in
\Theta, \overrightarrow{s}(\overrightarrow{\theta})=
(s^{1}(\overrightarrow{\theta}),...,s^{n}(\overrightarrow{\theta}))$.

With everyone weighed with weight $\alpha\{\cdot\}$ an average
geodetic $U _ {\alpha}$ with density (\ref {geod}) in \cite
{Chentsov, Moroz} two values of parameter communicate:

\begin{equation}
\overrightarrow{X}[\alpha]=\int_{C}\overrightarrow{x}\alpha\{d\overrightarrow{x}\}
 ,\, \label{aver}
\end{equation}

\begin{equation}
\overrightarrow{Y}[\alpha]: Y_{j}=y_{j}+M_{\overrightarrow{y}}[\ln
u_{\alpha}(\omega)-\ln
p(\omega;\overrightarrow{y})]\sum_{k}v_{jk}(\overrightarrow{y})
r^{k}(\omega;\overrightarrow{y}) ,\, \label{aver1}
\end{equation}
where $r^{k} (\omega; \overrightarrow {z})$ it is certain in (\ref
{r}) - (\ref {M}), a matrix $\overrightarrow {V} (\overrightarrow
{x}) = (\overrightarrow {v} _ {jk}) (\overrightarrow {x})$ is
return to an information matrix $\overrightarrow {W}
(\overrightarrow {x}) = (\overrightarrow {w^{jk}})
(\overrightarrow {x})$ (\ref {M1}). When $\overrightarrow {Y}
[\alpha] \in \Theta$, speak, that the law $P _ {\overrightarrow
{Y} [\alpha]}$ accompanies the weighed average geodetic $U _
{\alpha}$. The point $\overrightarrow {y} \in F \subset \Theta$ is
set in \cite {Chentsov} as the center of a cube

\begin{equation}
C_{r}=\{\overrightarrow{x}: |x_{j}-y_{j}|\leq r; j=1,...,n=dim
\overrightarrow{\Phi}\}  \label{kub}
\end{equation}
in space $\overrightarrow {y}$-local parameters of smooth compact
family
$\overrightarrow{\Phi}=\{P_{\overrightarrow{x}},\overrightarrow{x}\in
F \};  rn^{1/2}\leq\rho(\Phi)$; all cube (\ref {kub}) belongs
compact $K (\overrightarrow {\Phi}) \subset \Theta$, and on it
uniform estimations of derivatives (\ref {g}) - (\ref {constH})
are executed. The corresponding family $\{P_{\overrightarrow{x}},
\overrightarrow{x}\in C_{r}\}$ is designated in \cite {Chentsov}
$\overrightarrow{\Phi}(r)=\overrightarrow{\Phi}_{\overrightarrow{y}}(r)$
and refers to cubic.

In \cite {Chentsov, Moroz} it is proved, that

\begin{equation}
||\overrightarrow{Y}[\alpha]-\overrightarrow{X}[\alpha]||_{y}\leq
r^{2}H^{2}n^{3/2} ,\, \label{otsen}
\end{equation}
where $H$ is a constant from (\ref {constH}). For family
$\overrightarrow {\Phi}$ it is possible to specify the size $\rho
_ {0} (\overrightarrow {\Phi})$ such, that at $r <\rho _ {0}
(\overrightarrow {\Phi})$ the accompanying law $P _ {Y [\alpha]}$
exists, what were $\overrightarrow {y} \in F$ and a probabilistic
measure $\alpha\{\cdot\}$, and $Y [\alpha] \in C _ {2r} \subset K
(\overrightarrow {\Phi})$.

For information deviations (Kullback's entropy)

\begin{equation}
I[Q|P]=\int_{\Omega}[\frac{dP}{dQ}(\omega)\ln\frac{dP}{dQ}(\omega)]Q\{d\omega\}
 =-\int_{\Omega}[\ln \frac{dQ}{dP}(\omega)]P\{d\omega\}=
\int_{\Omega}[\ln \frac{dP}{dQ}(\omega)]P\{d\omega\} \label{Kulb}
\end{equation}
(last equality is fair, when laws $P$ and $Q$ in (\ref {Kulb}) are
mutually absolutely continuous) the correlation enters the name

\begin{equation}
I[P|U]=I[P|P_{Y}]+I[P_{Y}|U]+<\ln(dP/dP_{Y}),U-P_{Y}>
,\,\label{Kulb1}
\end{equation}
where $P _ {Y}$ is the accompanying law, $<f,P>=\int
f(\omega)P\{d\omega\}$, $U\in
\Gamma(N)=\Gamma(\overrightarrow{\Phi}_{N})$,
$\Gamma(\overrightarrow{\Phi}_{N})$ is an integrated convex cover
of initial family $\overrightarrow {\Phi}$ \cite {Chentsov},
family of laws with densities (\ref {geod}), containing a convex
cover of initial family $\overrightarrow {\Phi}$, $\overrightarrow
{\Phi}_{N}=\overrightarrow{\Phi}(r(N)),
r(N)=N^{-3/2}<\rho_{0}(\overrightarrow{\Phi})$.

In \cite {Chentsov, Moroz} the difference between $\ln u _
{\alpha} (\omega)$ and $\ln p (\omega; \overrightarrow {x})$ and
$u _ {\alpha} (\omega)$ and $p (\omega; \overrightarrow {x})$ is
estimated also. For cubic family $\overrightarrow {\Phi} (r)$ at
$r <\rho _ {0} (\overrightarrow {\Phi})$ the suspension $U _
{\alpha}$ is close to the accompanying law $P _ {Y [\alpha]}$:

\begin{equation}
|\ln u_{\alpha}(\omega)-\ln p(\omega;\overrightarrow{Y})|\leq
r^{2}[B_{2}+B_{3}h^{(y)}(\omega)] ,\, \label{otsenln}
\end{equation}

\begin{equation}
|u_{\alpha}(\omega)-p(\omega;\overrightarrow{Y})|\leq
r^{2}g^{(y)}[B_{2}+B_{3}h^{(y)}(\omega)]B_{4} ,\,
\label{otsendistr}
\end{equation}
where $B _ {2} =B _ {2} (n, H) =4nH, B _ {3} =B _ {3} (n, H)
=4n+Hn^{3/2}, B _ {4} = \exp [4n \rho _ {0} H]$, values $h^{(y)},
g^{(y)}, r, H$ are certain in (\ref {g}) - (\ref {constH}),
(\ref{kub}). In \cite {Chentsov} conditions of convergence of
distributions (\ref {geod}) and compactness of an integrated
convex environment of family $\overrightarrow {\Phi}$ are written
down also.

\section{Nonequilibrium Statistical Operator as the weighed average geodetic
of laws of family of quasi-equilibrium distributions}

In work \cite {Ryazanov01} logarithm of NSO $ \rho (t)$ \cite
{Zub71, Zub80, Zub99} is interpreted as averaging of the logarithm
of quasi-equilibrium distribution $\rho _ {q}$ \cite {Zub80,
Zub99} from various time arguments on distribution $p _ {q} (u)$
of lifetime of system (time of the first achievement of a level):

\begin{equation}
\ln \rho(t)=\int_{0}^{\infty}p_{q}(u)\ln \rho_{q}(t-u,-u)du
 ,\, \label{nso}
\end{equation}
where $u=t-t _ {0}$ is a random variable of a lifetime of system,
$t $ is a present situation of time, $t _ {0}$ is a random
variable of the initial moment of time, "birth"  of system. The
value $u=t-t _ {0}$ is equal also to the random moment of the
first achievement of a zero level \cite {Fel, Red} during the
moment $t _ {0}$ in return time, at $t \mapsto-t$, (\ref {time}).

\begin{equation}
\Gamma_{x}=inf\{t:y(t)=0\}, \quad y(0)=x>0.
 \label{time}
\end{equation}

 If $p _ {q} (u) = \varepsilon e^{\varepsilon u}$,
distribution $p _ {q} (u)$ has an exponent form with
$\varepsilon=1 / \langle \Gamma \rangle$, where $\langle \Gamma
\rangle = \langle t-t _ {0} \rangle$ is average a lifetime of
system, from (\ref {nso}) is received NSO in the form of Zubarev
\cite {Zub71, Zub80, Zub99}. Quasi-equilibrium distribution $\rho
_ {q}$ is equal \cite {Zub80}

\begin{equation}
\ln
\rho_{q}(t_{1},t_{2})=-\Phi(t_{1})-\sum_{n}F_{n}(t_{1})P_{n}(t_{2})
 ,\, \label{relev}
\end{equation}
where dependence $P _ {n} (t _ {2})$ is understood as realization
of laws of conservation \cite {Zub71} when operators $P _ {n}$ in
a quantum case are considered in Geyzenberg representation, and in
case of classical mechanics Geyzenberg representation is replaced
with action of the operator of evolution, for example

 \begin{equation}
H(x,t)=e^{-iLt}H(x); \quad
\rho_{q}(t-u,-u)=e^{-iuL}\rho_{q}(t-u,0)
 ,\, \label{evol}
\end{equation}
where $L$ is Liouville operator \cite {Zub71}. Values $P _ {n}$ in
(\ref {relev}) represent dynamic variables (for example, energy,
number of particles, etc.); their average values give a set of
observable values, $F _ {n}$ are Lagrange multipliers connected
with intensive thermodynamic variables (temperature, chemical
potential, etc.). Similar expressions enter the name not only for
hydrodynamical, but also for a kinetic stage of evolution of
system \cite {Zub80, Zub99}.

Expressions (\ref {relev}) for $\rho _ {q}$ correspond to
exponential family (\ref {exp1}) and coincide with it at

\begin{equation}
\rho_{q}(t;\omega)=p(\omega;\overrightarrow{\theta})/p_{0}(\omega);
\quad \overrightarrow{\theta}=\overrightarrow{x}=u=t-t_{0},
\label{sravn}
\end{equation}
$$
\Phi(t-u)=\Psi(\overrightarrow{s}(\overrightarrow{\theta})), \quad
P_{n}=q_{n}(\omega); \quad
F_{j}(t-u)=-s^{j}(\overrightarrow{\theta}) .
$$

 At performance of conditions (\ref
{sravn}) expression for NSO (\ref {nso}) coincides with (\ref
{geod}) at

\begin{equation}
\alpha\{d\overrightarrow{x}\}=p_{q}(u)du, \quad
u=\overrightarrow{x}=t-t_{0}=\Gamma, \quad H[\alpha]=0 .
\label{sravn1}
\end{equation}

Let's show, that for NSO (\ref {nso}) $H [\alpha] =0$. The value
$H [\alpha]$ in (\ref {geod}) - (\ref {norm}) for distribution
$p(\omega, \overrightarrow {x})$ a kind (\ref {exp1}) at
performance of conditions (\ref {sravn}) - (\ref {sravn1}) is
equal

\begin{equation}
H[\alpha]=\Psi(\int_{C}\alpha\{d\overrightarrow{x}\}
s^{j}(\overrightarrow{x})q_{j}(\omega))-
\int_{C}\alpha\{d\overrightarrow{x}\}\Psi(s^{j}(\overrightarrow{x})q_{j}(\omega))
,  \label{normnso}
\end{equation}
where $\Psi (s)$ it is certain in (\ref {norm1}). In work \cite
{Zub71} where Zubarev's NSO  $\rho _ {Z}$ corresponds to an
invariant part from the logarithm of the locally-equilibrium
operator $\rho _ {l}$ (or $\rho _ {q}$) \cite {Zub71}, i.e.

\begin{equation}
\ln \rho_{Z}(t)=\varepsilon\int_{0}^{\infty}e^{-\varepsilon u}\ln
\rho_{l}(t-u,-u)du
 ,\, \label{nsozub}
\end{equation}
for

$$
\Phi_{l}(t-u)=\Psi(s(t-u))=\ln Sp
\exp\{-\sum_{m}\int_{V}F_{m}(\overrightarrow{r},t-u)
P_{m}(\overrightarrow{r},-u)d\overrightarrow{r}\},
$$
where $\overrightarrow {r}$ is spatial coordinate, $V$ is the
volume of system, dependence $P _ {m}$ from $u$ is given in (\ref
{evol}), are written down correlations

$$
\Phi_{l}=\varepsilon\int_{0}^{\infty}e^{-\varepsilon
u}\Phi_{l}(t-u)du=\varepsilon\int_{0}^{\infty}due^{-\varepsilon
u}\ln Sp \exp\{-\sum_{m}\int_{V}F_{m}(\overrightarrow{r},t-u)
P_{m}(\overrightarrow{r},-u)d\overrightarrow{r}\};
$$

$$
\Phi_{l}=\ln Sp
\exp\{-\sum_{m}\varepsilon\int_{0}^{\infty}\int_{V}e^{-\varepsilon
u}F_{m}(\overrightarrow{r},t-u)
P_{m}(\overrightarrow{r},-u)dud\overrightarrow{r}\}.
$$
Substituting these expressions in (\ref {normnso}), we receive,
that $H [\alpha] =0$.

Expression (\ref {aver}) for case NSO when correlations (\ref
{sravn} ) - (\ref {sravn1}) are carried out, defines average
lifetime of system, and expression (\ref {aver1}) assess for it,
quasi-projection in terms \cite {Chentsov, Moroz} with accuracy
(\ref {otsen}). Expressions (\ref {r}) - (\ref {M}) coincide with
the operator of entropy production \cite {Zub71, Zub80, Zub99}
$\hat {\sigma} (t-u,-u) = \partial \ln \rho _ {q} (t-u,-u) /
\partial u$.

In work \cite {Ryazanov01} estimations of a kind (\ref {aver1})
for a cube (\ref {kub}) with the center in a point
$\overrightarrow {y} =0$

\begin{equation}
Y[\alpha]=\int_{\Omega}[\ln \rho(t)-\ln
\rho_{q}(t,o)]\frac{\hat{\sigma}(t,0)\rho_{q}(t,o)}{\langle\hat{\sigma}^{2}(t,0)\rangle_{q}}dz
,  \label{avernso}
\end{equation}
where $z = (q _ {1}..., q _ {N}; p _ {1}..., p _ {N})$ is set of
coordinates $q$ and impulses $p$ all particles of system, $z =
\omega$ in (\ref {g}) - (\ref {M1}); $\langle ... \rangle _ {q} =
\int _ {\Omega} ... \rho _ {q} (t, o) dz$, are compared to the
similar expressions received directly from NSO. In \cite
{Ryazanov01} the example of calculation of average a lifetime for
system of neutrons in a nuclear reactor is set.

In a correlation (\ref {avernso}) where $Y \sim \langle \Gamma
\rangle$, entropy production $\hat {\sigma}$ and entropy fluxes
\cite {Zub71, Zub80, Zub99} enters. And, at $\hat {\sigma}
\rightarrow 0, \langle \Gamma \rangle \sim \frac {0} {0^{2}}
\rightarrow \infty$, and at $\hat {\sigma} \rightarrow \infty,
\langle \Gamma \rangle \sim \frac {\infty} {\infty^{2}}
\rightarrow 0$. Thus, a lifetime of the system depends on entropy
production in system and entropy fluxes, from an exchange of
entropy between system and an environment.

 Integrating in parts expression
(\ref {nso}), we receive, that at $\int p _ {q} (u) du _ {| u=0}
=-1, \int p _ {q} (u) du _ {| u \rightarrow \infty} =0$, \cite
{Ryazanov01},

\begin{equation}
\ln \rho(t)=\ln \rho_{q}(t,0)-\int_{0}^{\infty}(\int p_{q}(u)du)
\hat{\sigma}(t-u,-u)du . \label{nso1}
\end{equation}

From here
$$
-\int_{0}^{\infty}(\int p_{q}(u)du)\hat{\sigma}(t-u,-u)du=\ln
\rho(t)-\ln \rho_{q}(t-Y,-Y)+ \ln \rho_{q}(t-Y,-Y)-\ln
\rho_{q}(t,0).
 $$

The first item in the right part of the received expression, value
$\ln \rho (t) - \ln \rho _ {q} (t-Y,-Y)$ we shall estimate by
means of a correlation (\ref {otsenln}) at performance of
conditions (\ref {sravn}) - (\ref {sravn1}), when $u _ {\alpha}
(\omega) = \rho (t), p (\omega, x) = \rho _ {q} (t-x,-x)$. The
second item is estimated by means of received in \cite {Chentsov}
correlation

$$
\ln [p(\omega;x)/p(\omega;Y)] \stackrel{<}{>}
 \pm||x-Y||h^{(y)}(\omega) \, ,
$$
where $h^{(y)}$ from (\ref{consh}). Then

$$
-\int_{0}^{\infty}(\int p_{q}(u)du)\hat{\sigma}(t-u,-u)du\leq
r^{2}[4nH+(4n+Hn^{3/2})h^{(y)}(\omega)]+\rho(\Phi)h^{(y)}(\omega).
$$
In the left part value $\hat {\sigma}$ it is estimated by means of
a correlation (\ref {consh}), and

$$
-\int_{0}^{\infty}(\int p_{q}(u)du)\hat{\sigma}(t-u,-u)du\leq
-h^{(y)}(\omega)\int_{0}^{\infty}(\int
p_{q}(u)du)du=h^{(y)}(\omega)\langle\Gamma\rangle .
$$

Thus, average a lifetime $\langle \Gamma \rangle$ is limited, and
in (\ref {nsozub}) $\varepsilon=1 / \langle \Gamma \rangle \neq 0$
though in \cite {Zub71, Zub80, Zub99} limiting transition
$\varepsilon \rightarrow 0$ after thermodynamic limiting
transition is spent.  The reason of it that in \cite {Chentsov,
Moroz} is considered cubic family in the limited cube (\ref {kub})
with lifetimes limited by the value $r$. In theory of NSO \cite
{Zub71, Zub80, Zub99} limiting transition $\varepsilon \rightarrow
0$ is carried out after thermodynamic limiting transition
$V\rightarrow \infty, N \rightarrow \infty, V/N=const$.
Intuitively clearly, that a lifetime  of infinite greater systems
will be infinitely great.

Similar estimations enter the name and for expression (\ref
{otsendistr}) in view of received from (\ref {nso1}) expression

$$
\rho(t)-\rho_{q}(t,0)=\sum_{k=1}^{\infty}[-\int_{0}^{\infty}du(\int
p_{q}(u)du)\hat{\sigma}(t-u,-u)du]^{k}\rho_{q}(t,0).
$$

\section{Superstatistics as the weighed average
geodetic of laws of family Gibbs distributions depending on
nonequilibrium parameter}

Distributions of a kind (\ref {geod}) describe not only NSO. In
works \cite {Beck, BeckCohen} are entered superstatistics of A
type, when

\begin{equation}
p(E)=B(E)/Z_{A}; \quad B(E)=\int_{0}^{\infty}f(\beta)\exp\{-\beta
E\}d\beta; \quad Z_{A}=\int_{0}^{\infty}B(E)\omega(E)dE,
\label{supA}
\end{equation}
where $f (\beta)$ is some distribution of value $\beta$, return
temperature, the intensive thermodynamic variable, the conjugate
of energy $E$, and B type, when

\begin{equation}
p(E)=\int_{0}^{\infty}f(\beta)\frac{\exp\{-\beta
E\}}{Z(\beta)}d\beta; \quad Z(\beta)=\int_{0}^{\infty}\exp\{-\beta
E\}\omega(E)dE; \quad  \int_{0}^{\infty}f(\beta)d\beta=1.
\label{supB}
\end{equation}

Expression (\ref {supA}) passes in (\ref {supB}) at replacement
$\tilde {f} (\beta) = \frac {cf (\beta)} {Z (\beta)}$, $c=const$.
The special case of superstatistics, at function $f (\beta)$, set
in the form of gamma-distribution, leads to Tsallis distributions
\cite {Tsal} with $\beta _ {0} = \int \beta f (\beta) d \beta$.

If in (\ref {supA}) instead of distribution
$p(\beta;E)=exp\{-\beta E\}/Z(\beta)$ to use distribution (\ref
{exp1}) with $\overrightarrow {x} = \overrightarrow {\theta}$ and
$p (\omega; \overrightarrow {x}) =p (\omega; \overrightarrow
{\theta})/p _ {0} (\omega) = p (\beta (\theta); E)$ for a case A
from (\ref {supA}), substituting (\ref {exp1}) in (\ref {geod}),
we receive concurrence with (\ref {supA}) at

\begin{equation}
s(\theta)=-\beta(\theta), \quad q=E; \quad \ln
(p(\omega;\overrightarrow{\theta})/p_{0}(\omega))=-\beta(\theta)E-\ln
Z(\beta(\theta)); \label{supaver}
\end{equation}
$$
u_{\alpha}(\omega)=p(E); \quad Z(\beta(\theta))=\int
\exp[-\beta(\theta)E]\omega(E)dE;
$$
$$
\int \ln
p(\beta(\theta);E)\alpha(\theta)d\theta=-\int\beta(\theta)\alpha(\theta)d\theta
E-\int \ln Z(\beta(\theta))\alpha(\theta)d\theta=
$$
$$
\ln \int_{0}^{\infty}f(\beta)\exp\{-\beta E\}d\beta=\ln B(E).
$$

In this case $H [\alpha] = \ln Z _ {A} \neq 0$, unlike (\ref
{sravn1}). For value $H [\alpha]$ in \cite {Chentsov, Moroz} are
lead an estimation:

$$
0\leq -H[\alpha]\leq 4nr^{2}H
$$
in \cite{Chentsov}, and

$$
-H[\alpha]\leq \int_{C}I[P_{x}|R]\alpha\{dx\}
$$
in \cite {Moroz} where existence of such distribution of
probabilities $R(\cdot)$ on $(\Omega, \overrightarrow {S})$, is
supposed, that $sup_{C}I[P_{x}|R]<\infty$ (designations
$C,(\Omega, \overrightarrow {S})$ correspond (\ref {geod})). From
(\ref {supaver}) communication of functions of distribution
$f(\beta)$ and $\alpha(\theta)$ is defined.

 The parameter $\theta$ in (\ref {supaver})
represents some extensive thermodynamic parameter corresponding
internal thermodynamic parameter, describing nonequilibrium of the
system \cite {Leon}. It can be coordinate of the center of weights
in a field of weight, the electric moment of dielectric in an
external electric field \cite {Leon}, number of phase jumps in
problems of phase synchronization \cite {Strat61, Tich}, etc.
Average values are equal

$$
<\theta>=\int \theta\alpha(\theta)d\theta; \quad
\beta_{\theta}=\int \beta(\theta)\alpha(\theta)d\theta .
$$

Generally $\beta _ {\theta}$ does not coincide with $\beta _ {0}$.
But for greater systems $\beta _ {\theta} \sim \beta _ {0}$. As
$\overrightarrow {\theta}$ is vector value expressions of a kind
(\ref {supaver}), received from (\ref {exp1}), (\ref {geod}),
describe also superstatistics with several fluctuating
thermodynamic parameters. Such expressions are received in work
\cite {Ryazanov05,Ryazanov06}.

The estimations
lead in section 3 for NSO, applicable and for superstatistics. So,
for superstatistics A type the correlation (\ref {otsenln}) enters
the name in the form of (at $n=1$)

$$
\ln (\int_{0}^{\infty}f(\beta)e^{-\beta
E}d\beta)+\beta(\theta[\alpha])E+\ln Z(\beta(\theta[\alpha]))-\ln
\int_{0}^{\infty}(\int_{0}^{\infty}f(\beta) e^{-\beta
E}d\beta)\omega(E)dE\leq
$$
$$
r^{2}[4H+(4+H)h^{(y)}(\omega)],
$$
where ( (\ref{consh}), (\ref{constH}), (\ref{aver1}))

$$
\frac{\partial\ln
p(\omega;\theta)}{\partial\theta}=\frac{\partial\ln
p(\omega;\theta)}{\partial\beta(\theta)}\frac{\partial\beta(\theta)}
{\partial\theta}=\frac{\partial\beta(\theta)}{\partial\theta}[\langle
E(\beta(\theta))\rangle-E]\leq h^{(y)}(\omega); \quad  \langle
E(\beta(\theta))\rangle=-\frac{\partial\ln
Z(\beta(\theta))}{\partial\beta(\theta)};
$$

$$
\int (\frac{\partial\beta(\theta)}{\partial\theta})^{4}[\langle
E(\beta(\theta))\rangle-E]^{4}
\frac{e^{-\beta(\theta)E}}{Z(\beta(\theta))}\omega(E)dE\leq
M_{\theta}[h^{(y)}(\omega)]^{4}\leq H^{4};
$$

$$
\theta[\alpha]=y+\int \ln B(E)\frac{[\langle
E(\beta(y))\rangle-E]}{\frac{\partial \beta(y)}{\partial y}
\frac{\partial^{2}\ln
Z(\beta(y))}{\partial\beta^{2}(y)}}\frac{e^{-\beta(y)E}}{Z(\beta(y))}\omega(E)dE-
\beta(y)\frac{1}{\frac{\partial \beta(y)}{\partial y}}.
$$

If the center of a cube $C _ {r}$ (\ref {kub}) is located in a
point $y=0$, $\theta \leq r$. To estimate an arrangement of the
center of a cube and size of parameter $r$, it is necessary to
know the physical nature of parameter $\theta$, spending
consideration for a concrete physical situation.

Expression (\ref
{otsendistr}) for superstatistics of A type and the correlations
(\ref {supA}) becomes

$$
|\frac{B(E)}{Z_{A}}-\frac{e^{-\beta(\theta[\alpha])E}}{Z(\beta(\theta[\alpha]))}|\leq
r^{2}g^{(y)}(\omega)[4nH+h^{(y)}(\omega)(4n+Hn^{3/2})]
e^{4n\rho_{0}H}.
$$

The value $ \rho _ {0} $, certain before expression (\ref {Kulb}),
too it is possible to estimate after the obvious task of parameter
$\theta$.

\section{Conclusion}

In works \cite {Chentsov, Moroz} a number of results important for
statistical physics contains. So, results of sections 3-5 are
formulated also by means of developed in \cite {Chentsov, Moroz}
projective methods which importance is emphasized in theory of NSO
\cite {Zub71, Zub80, Zub99}. The problem A of projecting \cite
{Chentsov, Moroz} corresponds to a finding of a minimum of
Kullback's entropy (\ref {Kulb}) in nonequilibrium system \cite
{Strat82}, \S 29.5. Expressions for divergence of Amari, Kagan,
Csiszar \cite {Moroz} also are compared with entropy functionals,
used in Tsallis statistics \cite {Tsal} (for example, to
information quantities of Renyi). In \cite {Chentsov, Moroz} the
methods allowing strictly to approach to important for NSO the
problem of selection of basic variables of quasi-equilibrium
distribution \cite {Zub71, Zub80, Zub99} are developed.
Interesting the problem of a finding of interpretation in the
statistical physics of such concepts as statistical decision
rules, risk, asymmetrical pythagorean geometry \cite {Chentsov} is
represented.

\end{document}